\documentclass[universe,review,accept,oneauthor,pdftex]{Definitions/mdpi} 
\firstpage{1} 
\pubvolume{9}
\issuenum{9}
\articlenumber{402}
\pubyear{2023}
\copyrightyear{2023}
\externaleditor{Academic Editor: Nicolas Chamel}
\datereceived{4 August 2023 } 
\dateaccepted{30 August 2023 } 
\datepublished{1 September 2023} 
\hreflink{https://doi.org/10.3390/\linebreak universe9090402} 
\pdfoutput=1

\usepackage[T1]{fontenc}
\usepackage[utf8]{inputenc}

\usepackage{bm}
\usepackage{comment}
\usepackage{booktabs}
\usepackage{hyperref}
\usepackage{natbib}
\usepackage{graphicx}

\usepackage{color}

\usepackage{hyperref}
\makeatletter
\def\UrlAlphabet{%
      \do\a\do\b\do\c\do\d\do\e\do\f\do\g\do\h\do\i\do\j%
      \do\k\do\l\do\m\do\n\do\o\do\p\do\q\do\r\do\s\do\t%
      \do\u\do\v\do\w\do\x\do\y\do\z\do\A\do\B\do\C\do\D%
      \do\E\do\F\do\G\do\H\do\I\do\J\do\K\do\L\do\M\do\N%
      \do\O\do\P\do\Q\do\R\do\S\do\T\do\U\do\V\do\W\do\X%
      \do\Y\do\Z}
\def\UrlDigits{\do\1\do\2\do\3\do\4\do\5\do\6\do\7\do\8\do\9\do\0}
\g@addto@macro{\UrlBreaks}{\UrlOrds}
\g@addto@macro{\UrlBreaks}{\UrlAlphabet}
\g@addto@macro{\UrlBreaks}{\UrlDigits}
\makeatother

\newcommand\g {$\gamma$}

\newcommand{\apj}{ Astrophys. J.}			
\newcommand{\apss}{Ap\&SS}		
\newcommand{\prd}{Phys.~Rev.~D}		
\newcommand{\prl}{Phys.~Rev.~Lett.}	
\newcommand{\ssr}{Space~Sci.~Rev.}	
\newcommand{\nat}{Nature}		

\newcommand\Msun{M_\odot}
\newcommand\rs[1]{_\mathrm{#1}}
\newcommand\tch{t\rs{ch}}
\newcommand\Esn{E\rs{sn}}
\newcommand\Mej{M\rs{ej}}
\newcommand\vpsr{v\rs{psr}}
\newcommand\rhoism{\rho\rs{ism}}
\newcommand\Edot{\dot{E}}
\newcommand\Empd{E\rs{MPD}}
\newcommand\gmpd{\gamma\rs{MPD}}

\Title{Evolved Pulsar Wind Nebulae}
\TitleCitation{Evolved Pulsar Wind Nebulae}

\Author{Barbara 
 Olmi $^{1,2}$\orcidB{}}

\AuthorNames{Barbara Olmi}

\AuthorCitation{Olmi, B.}

\address{$^{1}$ \quad INAF---Osservatorio Astronomico di Palermo, Piazza del Parlamento 1, 90134 Palermo, Italy; barbara.olmi@inaf.it\\
$^{2}$ \quad INAF---Osservatorio Astrofisico di Arcetri, Largo E. Fermi, 5, 50125 Firenze, Italy}

\abstract{Based on the expected population of core collapse
supernova remnants and the huge number of detected pulsars in the Galaxy, still representing only a fraction of the real population, pulsar wind nebulae are likely to constitute one of the largest classes of {extended} Galactic sources in many energy bands. For simple evolutionary reasons, the majority of the population is made of evolved systems, whose detection and identification are complicated by their reduced luminosity, the possible lack of X-ray emission (that fades progressively away with the age of the pulsar), and by their modified morphology with respect to young systems. Nevertheless they have gained renewed attention in recent years, following the detection of misaligned X-ray tails protruding from an increasing number of nebulae created by fast moving pulsars, and of extended TeV halos surrounding aged systems. Both these features are clear signs of an efficient escape of particles, with energy close to
the maximum acceleration limit of the pulsar. 
Here we discuss the properties of those evolved systems and what we have understood about the process of particle escape, and the formation of observed features.}

\begin{document}
\section{Introduction}
\label{sec:intro}
\begin{figure}[H]
	\includegraphics[width=.73\textwidth]{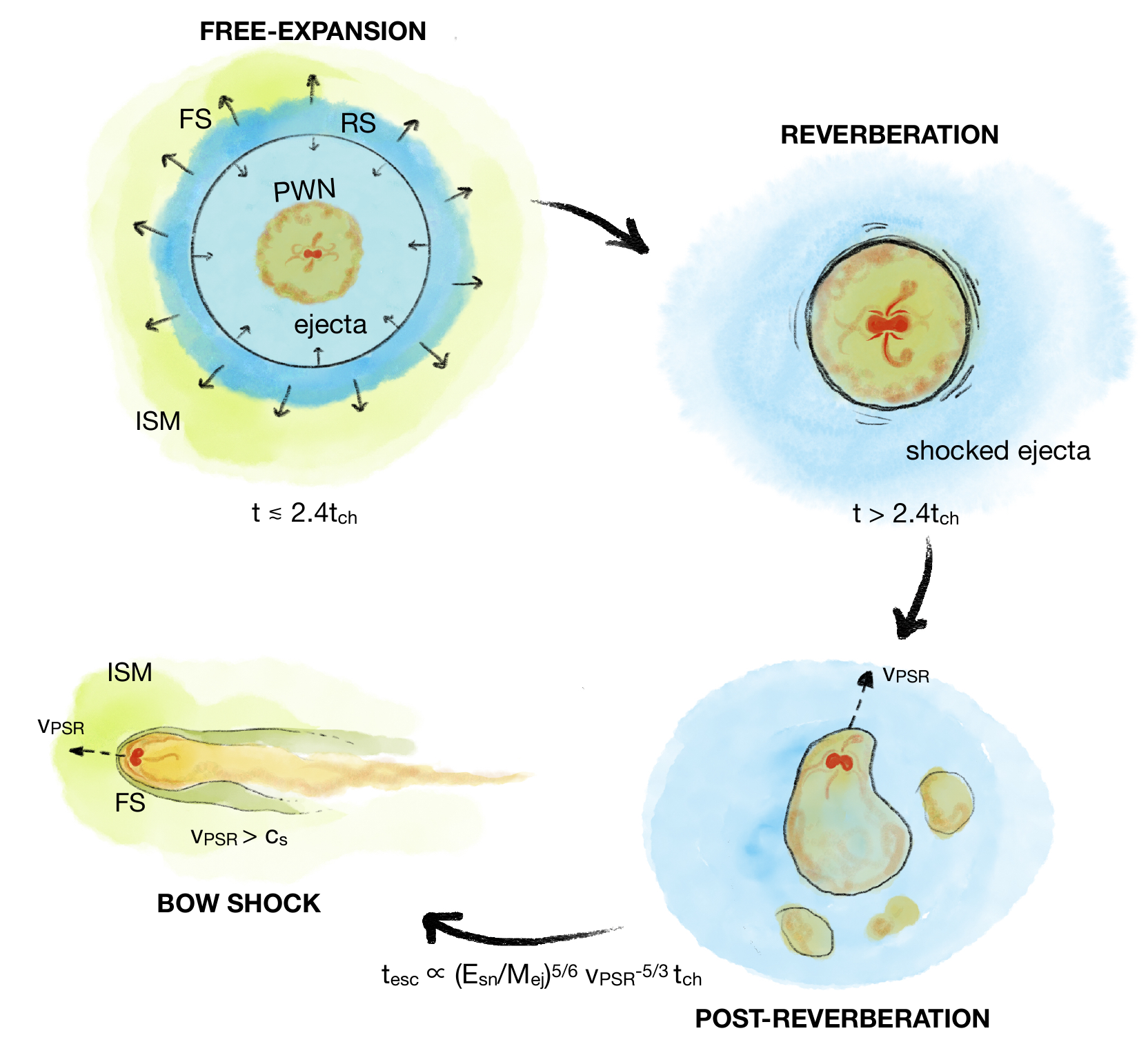}
    \caption{Illustration of the main phases of the evolution of a PWN. The~third stage is sub-divided into two different steps: post-reverberation and bow shock phase. The~typical duration of the first stage is given by the implosion time of the supernova reverse shock (RS), in~terms of the characteristic time of the system ($\tch$), defined in the main text. The~duration of the second phase, on~the other hand, depends on the specific parameters of the system and might range between 5 and 50 $\tch$. The~very last stage starts when the pulsar emerges out of the supernova remnant bubble, at~a time roughly given by $t\rs{esc}\propto(\Esn/\Mej)^{5/6}\vpsr^{-5/3}\,\tch\simeq 150 \, (\Mej/\Msun)^{-5/6} \, \tch$, for~a typical pulsar speed of 350 km/s. Once escaped, the~PWN suddenly becomes a bow shock PWN, since the pulsar velocity is generally much larger than the sound speed in the ambient medium ($c_s$).}
    \label{fig:pwnsketch}
\end{figure}

The evolution of pulsar wind nebulae (PWNe hereafter) can be roughly divided into three main stages, characterized by a different interaction between the PWN and its surroundings.
{ A sketch illustrating these phases can be seen in Figure~\ref{fig:pwnsketch}}.

At the beginning, the~PWN is in \textbf{free-expansion}: 
it expands with mild acceleration in the un-shocked and cold ejecta of the supernova remnant (SNR). In~this phase, the PWN behaves as if it were isolated: there is no direct interaction with the SNR, except~for the obvious fact that the ejecta contains the nebula, with~the formation of a contact discontinuity (CD) separating the two \citep{Reynolds_Chevalier:1984}.
How long this phase last essentially depends on three parameters: the initial spin-down luminosity of the pulsar, $L_0$, its spin-down time, $\tau_0$, and~the supernova (SN) explosion energy, $\Esn$. The~combination of them ($\lambda_E=L_0\tau_0/\Esn$) gives an indication of the energetics of the PWN related to the SNR. The~larger is $\lambda_E$, the~shorter is the free-expansion phase, since the PWN expands faster and, consequently, it enters into contact with the reverse shock (RS) of the SN sooner \citep {Bandiera:2023}.
On the contrary, low energetic systems expands very slowly, and~their free-expansion phase can last for a much longer time. 
The limit is the RS implosion time, that can be estimated as $\sim$$2.4\, (\Esn^{-1/2}\,\Mej^{5/6}\,\rhoism^{-1/3})$, where $\Mej$ is the mass of the SNR ejecta and $\rhoism$ is the mass density of the interstellar \mbox{medium \citep{T&M99,Bandiera:2021}.} The~quantity in parentheses represents the so-called characteristic time ($\tch$) of the SNR, that is a perfect scaling for the evolutionary phases of the PWN + SNR system~\citep{T&M99,Bandiera:2020}.

The following stage starts when the RS reaches the PWN boundary. This phase is known as ``\textbf{reveberation}'', due to the prediction by one-zone and 1D models of a number of oscillations of the PWN between compressions and re-expansions under the pressure exerted by the SNR (see, e.g.,~\citet{Olmi_Bucciantini:2023} for a detailed discussion and a complete list of references).
During this phase, the~PWN may experience important morphological and spectral modifications. The~interaction with the RS, almost certainly affected by some degree of anisotropy, can lead to very asymmetric PWNe \citep{Kolb:2017}. 
Depending on the intensity of the compression, the~PWN can also substantially change its spectral behavior. Strong compression produces an increase in the magnetic field which, in~turn, causes rapid burning of the particles, at~progressively lower energies the more the magnetic field increases \citep{Torres_supereff:2018,Bandiera:2023b}.
The outcome of this second phase is therefore difficult to predict a~priori.

Evolved PWNe can be very faint and extended, if~the pulsar has a slow proper motion and it remains close to its birth place for its entire evolution. 
However, the pulsar population is characterized by a large average kick velocity ($\sim$350 km/s \citep{FGK:2006,Sartore:2010,Verbunt:2017}), which likely causes a pulsar to leave its parent SNR bubble at some point in its evolution, as~the SNR expansion is effectively decelerated.
After having left the SNR, the~pulsar interacts directly with the external medium: the interstellar medium (ISM) in most cases.
Here the pulsar speed ($\vpsr$) is larger then the sound speed, and~the pulsar motion becomes supersonic (this in some cases already happens inside the SNR \citep{Gaensler_Slane:2006}). 
The supersonic motion induces the formation of a bow shock around the pulsar and its nebula that drastically reshapes  the PWN \citep{Wilkin:1996, Bucciantini_bowsI:2001}. These systems are called \textbf{bow shock pulsar wind nebulae} (BSPWNe hereafter) and~are characterized by an elongated cometary-like shape, with~the pulsar at the head of a long tail developed in the direction opposite to the pulsar motion, fed by the shocked pulsar wind.
This new morphology is governed by the balance of the ram-pressure of the pulsar wind with that of the ISM, seen as an incoming flow in the reference frame of the pulsar. This actually determines the thickness of the bow shock front, known as the stand-off distance, and~it represents the characteristic dimension for a BSPWN:
\begin{equation}\label{eq:standoff}
d_0 = \sqrt{\frac{\Edot(t)}{4\pi c \, \rhoism \vpsr^2}}\,,
\end{equation}
where $\Edot(t)$ is the spin-down power of the pulsar at a generic moment $t$.

To date, we have observed around 110 PWNe in different stages, plus $\sim$20 other candidate sources \citep{Olmi_Bucciantini:2023}.
However the number of PWNe in the Galaxy is estimated to be much larger; { a rough evaluation can be done} considering a rate of birth of pulsars in the Galaxy of 1/100 yr \citep{FGK:2006} and 
{100 kyr as the order of magnitude for the lifetime of a PWN at gamma-rays (the longer living energy band), obtaining an expected number of \mbox{1000 PWNe.} }
%
 The multi-wavelength identification of PWNe is rather difficult; the extended radio emission, especially in the case of evolved systems, might be contaminated by the background, and~most of the present interferometers lack in sensitivity to the required large angular scales.
On the other side, the~very low spatial resolution of present gamma-ray instruments makes the morphological identification of extended~sources very difficult.

Although the vast majority of the PWNe in the Galaxy are middle-aged or old, the~main detection band then remains X-rays, since there are not so many other extended sources in the Galaxy contributing at these energies.
For this reason, the attention of the scientific community in the past has focused mainly on the free-expansion phase,
%
%
but the lifetime of electrons able to emit in the X-ray band via synchrotron radiation in the nebular magnetic field is rather short \citep{Olmi_Bucciantini:2023},
\begin{equation}\label{eq:tau_sync}
\tau\rs{sync} \simeq 55.2 \;\left( \frac{B_{\upmu\mathrm{G} }}{100}\right)  \, \left( \frac{E\rs{ph, keV} }{1} \right) \; \mathrm{yr}\,,
\end{equation}
and, as~soon as the energy input from the pulsar drops below the threshold necessary to accelerate those particles, the~PWN fades rapidly away from this observational band. 
%
%
%
The perfect target of the initial evolutionary phase of PWNe is the Crab nebula, the~best studied object in the class, and~possibly one of the most studied in the entire Galaxy. It has been used as a benchmark for most of the models, despite it being a rather unique object, even among PWNe \citep{Amato_Olmi:2021}.
To date, we have very advanced relativistic magneto-hydrodynamic (MHD) numerical models that reproduce many of its observed properties, down to the finer details of the inner nebula that we observed thanks to the Chandra telescope. In~particular, we have understood that magnetic dissipation is so efficient when the magnetic field is appropriately described in 3D, that to reproduce the observed properties of the Crab, the~initial magnetization of the pulsar wind can easily exceed unity \citep{Porth:2014, Olmi:2016}. 

On the other hand, only a handful of middle-aged, reverberating systems have been identified to date (among which are Vela-X \citep{Blondin:2001}, the~Boomerang \citep{Kothes_boomerang:2006}, and the Snail \citep{Temim:2009,Temim:2015}). Their morphology is very asymmetric due to the ongoing interaction with the RS of the SNR, and~then a 3D description is mandatory to catch their {global} properties{ while, as~already seen for young systems \citep{Komissarov:2004,del-Zanna:2006,Camus:2009,Olmi:2015,Porth:2014}, 2D and 2.5D models efficiently catch a number of properties, especially in the inner nebula \citep{Temim:2015,Kolb:2017,Fateeva:2023}}. Due to the huge amount of time and the numerical resources the long evolution needed to reach this stage requires, a~very limited study has been performed to date, in~the hydrodynamic (HD) regime and dedicated to a specific source (the Snail, \citep{Kolb:2017}).
The principal difficulty with middle-aged objects is that their properties strongly depend on the past evolution, and~thus one cannot avoid to model the entire life of the system to correctly catch its present physical conditions.

The case of BSPWNe, despite that they are in general in a later evolutionary stage than the previous, is somehow simpler. 
Indeed, one can start modeling the system after it has left the SNR, since its dynamics from that point on are reshaped by interaction with the environmental medium, and~knowledge of past evolution is not so~important.

In this manuscript, we will mainly focus on this particular evolutionary phase. Although~interesting in itself from an evolutionary point of view, the~number of observations indicating that these sources are associated with the release of a large number of particles into the ambient medium has increased in recent years \citep{Olmi_Bucciantini:2023}, renewing the scientific community's interest in these particular PWNe.
Understanding with what efficiency and properties particles are released into the environment is particularly relevant to asses the role of pulsars as cosmic ray (CR) sources in the Galaxy, and~their relevance for the measured positron excess in the CR spectrum on Earth \citep{Adriani:2009,Aguilar:2013}.
Moreover, as~we will better discuss in the following, understanding how particles escape and behave in the source surroundings can give us fundamental information about the possibility for particles to confine themselves through self-developed instabilities that modify the properties of the ambient medium and, more in general, how  CR transport is modified in the vicinity of~sources.

This manuscript is structured as follows: in Section~\ref{sec:bs}, we recall the main observed properties of BSPWNe and, in~Section~\ref{sec:modbs}, how they have been modeled; in Section~\ref{sec:esc}, we describe the mechanism through which particles can escape those sources, especially from the bow shock front; then, in~Section~\ref{sec:out}, we discuss what happens to particles once they are outside the BSPWNe and how they arrange themselves to produce the oserved features. In~Section~\ref{sec:concl}, we draw our conclusions.
%
\section{Bow Shock Pulsar Wind~Nebulae}
\label{sec:bs}

Bow shock PWNe have been identified through the combination of observations in different bands, especially radio, H$_\alpha$, if the ambient medium is ionized, and~X-rays. To~date, we have counted 25 BSPWNe (see Table~2 in \citet{Olmi_Bucciantini:2023} for an updated list).
Of them, only a few (e.g., the Lighthouse \citep{Pavan:2014} and the Mouse \citep{Klingler:2018}) show an extended X-ray tail, and~are in fact associated with energetic pulsars ($\Edot=1.4\times 10^{36}$ erg/s and \mbox{$\Edot=2.5\times 10^{36}$ erg/s}, respectively), and still able to inject particles with enough energy to survive far away from the injection location.
Most of the BSPWNe are rather associated with faint pulsars, with~spin-down powers from one to three orders of magnitude lower than those previously mentioned (e.g., the Guitar Nebula has $\Edot=1.2\times 10^{33}$ erg/s \citep{de-Vries:2022}).
The fact that in general we can only detect a very limited tail extending behind the pulsar makes it difficult to reveal its spectral properties, with~a softening in the X-ray spectrum indicating the synchrotron cooling only determined for a few systems.
In some cases, we have also observed BSPWNe at radio energies, showing possibly even more extended tails  than at X-rays and, less frequently, in~the UV, in~association with the H$_\alpha$ emission probably produced by the heated ISM at the bow shock boundary.
No gamma-ray emission has been detected yet from BSPWNe, at~least not directly coming from the nebula, possibly due to the combination of the generally fainter pulsar luminosity if compared with young PWNe, and~to the lack of spatial resolution that affects present instruments, not able to resolve the tiny extension of the BSPWN head.

Around fifteen years ago, a~puzzling feature was detected in the surrounding of the Guitar nebula \citep{Hui_Becker:2007}: an extended (for $\sim$1 pc) X-ray feature, similar to a jet, apparently originating very close to the pulsar location at the head of the BPWNe and strongly misaligned with respect to the direction of motion of the pulsar (by $\sim$118$^{\circ}$), collimated for its entire length with no evidence for bending.
It was clear that, given its properties, that feature must be located outside the bow shock of the BSPWN.
Its appearance was so puzzling that it was initially thought it might be produced by an unidentified nearby \mbox{X-ray source \citep{Kargaltsev_Pavlov:2006}}.

The first theoretical interpretation of this feature came a couple of years later. 
In \citet{Bandiera:2008}, it is interpreted as being produced by energetic particles escaping from the bow shock and interacting with the ordered ambient magnetic field to emit synchrotron radiation.
For this process to work, the~escaping particles must have an energy close to the maximum theoretically available in the system, i.e.,~the energy related to the acceleration of particles by the maximum potential drop of the pulsar: $\Empd=e (\Edot/c)^{1/2}$. This corresponds to a maximum Lorentz factor:
\begin{equation}\label{eq:maxpd}
\gmpd=\frac{e}{m c^2} \sqrt{\frac{\Edot}{c}}\simeq 3\times 10^8 \,\left(\frac{\Edot}{10^{34} \, \mathrm{erg/s}}\right)\,.
\end{equation}
The 
 main problem with the interpretation is that, in~order for this particles to produce the observe X-ray emission, a~magnetic field of 40--50 $\upmu$G is required, namely about a factor of 10 larger than the one expected in the unperturbed~ISM.

\begin{figure}[H]
	\includegraphics[width=0.74\textwidth]{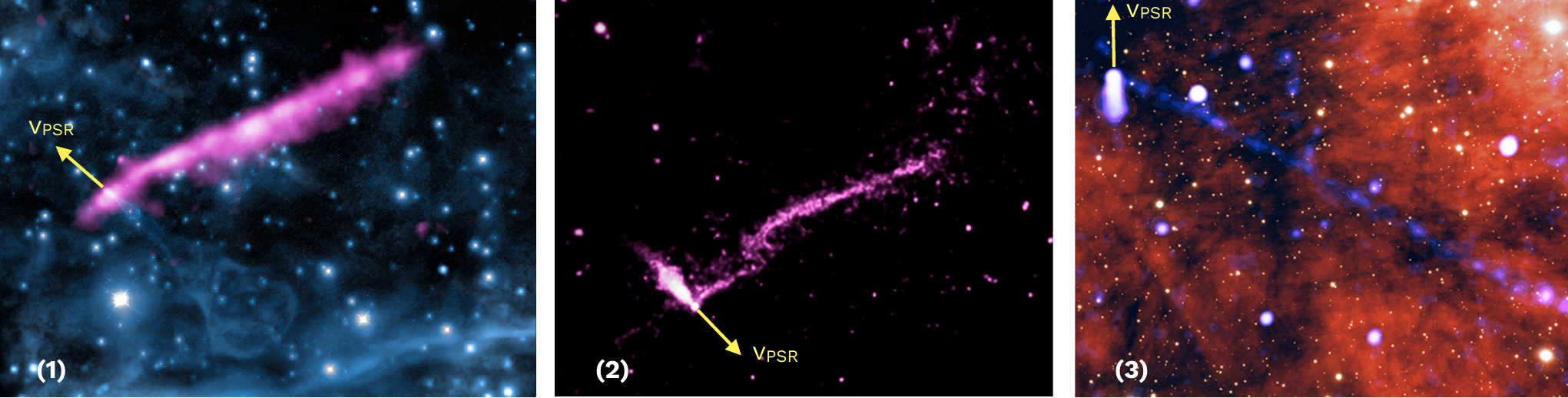}
    \caption{\textls[-15]{The most famous and representative examples of a BSPWN with its elongated misaligned X-ray feature. From~left to right: the Guitar nebula (\textbf{1}), the~Lighthouse nebula (\textbf{2}), and the BSPW associated to J2040 + 4415 (\textbf{3}).
    In all cases, we show the pulsar direction of motion with a yellow arrow, to~better appreciate the extreme inclination of the features.} Credits: (\textbf{1}) X-ray: NASA/CXC/UMass/S.Johnson~et~al. Optical: NASA/STScI \& Palomar Observatory 5-m Hale Telescope---(\textbf{2}) NASA/CXC/ISDC/L. Pavan~et~al.\linebreak   (\textbf{3}) NASA/CXC/Stanford University/M. de Vries/NSF/AURA/Gemini Consortium.} 
    \label{fig:bspwne_obs}
\end{figure}

In the following years, we detected an increasing number of similar features surrounding other BSPWNe \citep{Olmi_Bucciantini:2023}. 
Today, therefore, these features appear to be fairly common properties of evolved PWNe associated with fast-moving pulsars, and~the limited number of sources observed so far is perhaps due only to the fact that these features are very thin and faint, and~deep pointing is required to reveal them.
Aside from the Guitar nebula (see \citet{deVries_Romani:2022} for a recent review), the~better detailed features, showing similar properties, are those of the Lighthouse nebula \citep{Pavan:2014} and J2030 + 4415 \citep{deVries_Romani:2020}.
These fascinating three sources are all visible in Figure~\ref{fig:bspwne_obs}, together with their misaligned X-ray~tails.

A different evidence for the efficient escape of energetic particles from evolved pulsars is the observation of extended TeV halos, initially discovered around Geminga and Monogem \citep{Abeysekara:2017}.
These have been interpreted as being produced by the radiation emitted in the inverse Compton scattering process between escaped leptons of 10--100 TeV energies with the local photon fields.
In the following, other candidate TeV halos have been identified, as~well as extended PeV sources, possibly associated with pulsars \citep{LHAASO_EHE, LHAASOcat:2023}.

\section{Modeling Bow Shock~Nebulae}
\label{sec:modbs}
\textls[-15]{After a few initial attempts to model analytically (or semi-analytically) \mbox{BSPWNe \citep{Bandiera:1993, Wilkin:1996},}} at~the beginning of the millennium the first 2D HD numerical simulations proved the formation of the bow shock around a pulsar moving with a high proper motion in the ambient medium, and~the consequent development of the cometary-like PWN \citep{Toropina:2001, Bucciantini_BowsII_2002, van_der_Swaluw:2003}.

BSPWNe models were then extended in the following years to the MHD regime, showing that the magnetized wind does not modify the external layer of the shocked ISM, confirming the global morphology for the bow shock predicted by previous HD models \citep{Bucciantini:2005}.
The main limitation of these models is the imposed axi-symmetry, that on one side forces one to assume the pulsar spin-axis  to be aligned with the pulsar kick velocity (but there is no clue for this in observations  \citep{Johnston:2005,Johnston:2007,Ng_Romani:2007,Noutsos:2012,Noutsos:2013}) and on the other side limits the possible initial magnetization in the wind, not to generate artifacts on the axis that destroy the system geometry.
Moreover, while the importance of 3D modeling has been shown to be fundamental even in young PWNe to catch the correct geometry of the magnetic field  { at large scales} and the development of magnetic dissipation \citep{Porth:2014,Olmi:2016}, a~3D representation is the only possibility to investigate different inclinations of the pulsar spin-axis, magnetic axis and, direction of motion.
The first extension to a full 3D modeling was done in the non-relativistic HD regime by  \citet{Vigelius:2007}, considering both different latitudes in the pulsar wind and density gradients in the ambient medium. 
They show that a variation of the anisotropy in the pulsar wind does not reflect into a modification of the BSPWN head, while modifications in the ambient medium are fundamental in shaping the bow shock.
This result was then confirmed by following works, showing that the presence of inhomogeneity in the ambient medium deeply affects the morphology of a BSPWN \citep{Toropina:2019}, and~that very distorted tails (as the famous case of the Guitar nebula) can arise from the contamination of neutral atoms from the ISM  \citep{Morlino:2015, Olmi_Bucciantini_Morlino:2018}.

\begin{figure}[H]
    \centering
	\includegraphics[width=.4\textwidth]{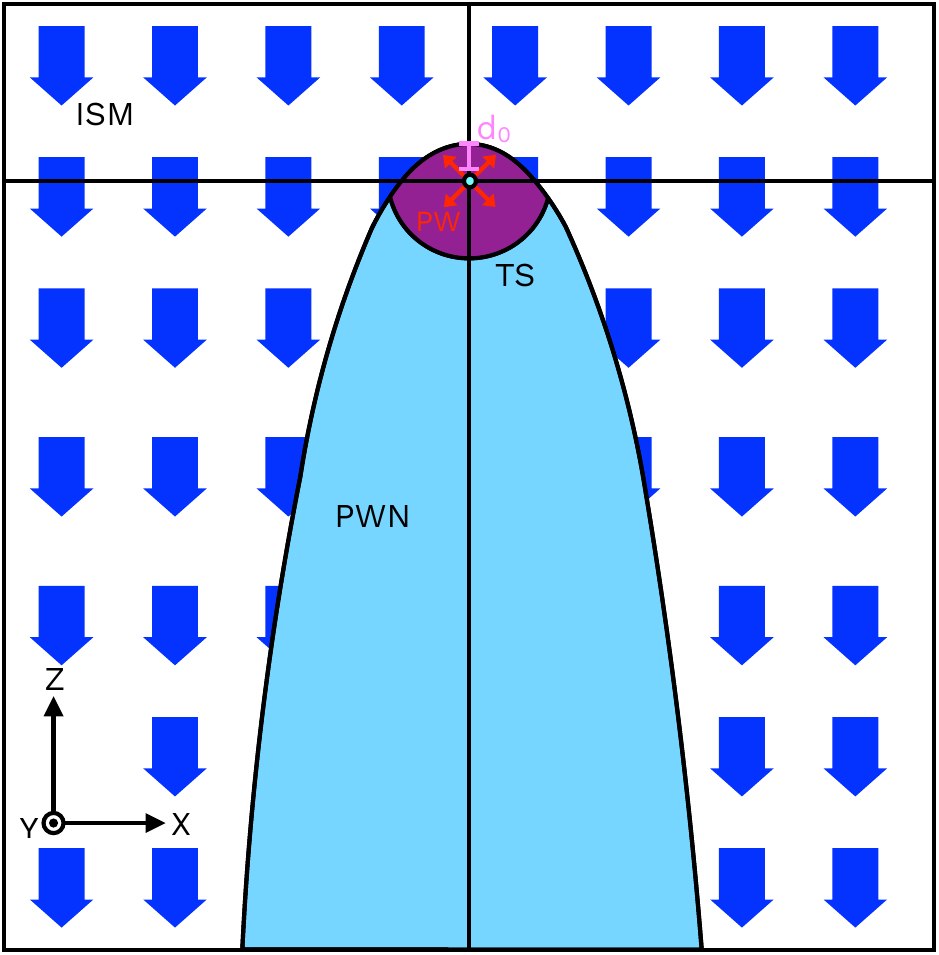}
    \caption{Sketch of the initial configuration for a 3D MHD simulation of a BSPWN. The~pulsar is located at the center of the cartesian grid, and~the axes directions are shown in the bottom-left triade. The~pulsar injects its outflow inside the termination shock (TS), whose shape is modified by the presence of the bow shock. The~ambient medium is seen by the pulsar (and nebula) as an incoming flow, in~this case directed along -Z. This configuration is the same used in \citet{Olmi_Bucciantini_2019_1}, where the initial structure of the bow shock was implemented using the analytic solution presented in \citet{Wilkin:1996}. The~stand-off distance introduced in Equation~(\ref{eq:standoff}) is shown at the bow shock head front (in pink color).} 
    \label{fig:bspwn_sketch} 
\end{figure}

Full 3D relativistic MHD numerical models have only become available in very recent \mbox{years \citep{Barkov:2019, Olmi_Bucciantini_2019_1}}. All these models do not treat the transitional phase when the pulsar is escaping the SNR, but~directly consider it in interaction with the ISM. 
The most convenient way to model the system is to move into the star reference frame; here, the pulsar is at rest at a specific position of the computational domain and~sees the ambient medium as an incoming cold flow, possibly magnetized, moving with a velocity $-\vpsr$. 
As for the pulsar injection, it is modeled similarly to what usually done for young systems \citep{Porth:2014,Olmi:2016}, with~the parameters defining the pulsar wind properties being the magnetization ($\sigma$), the~spin-to-magnetic field inclination, and the~anisotropy.
A sketch of a possible initial configuration is shown in Figure~\ref{fig:bspwn_sketch}.
%

%

\begin{figure}[H]
\centering
	\includegraphics[width=.6\textwidth]{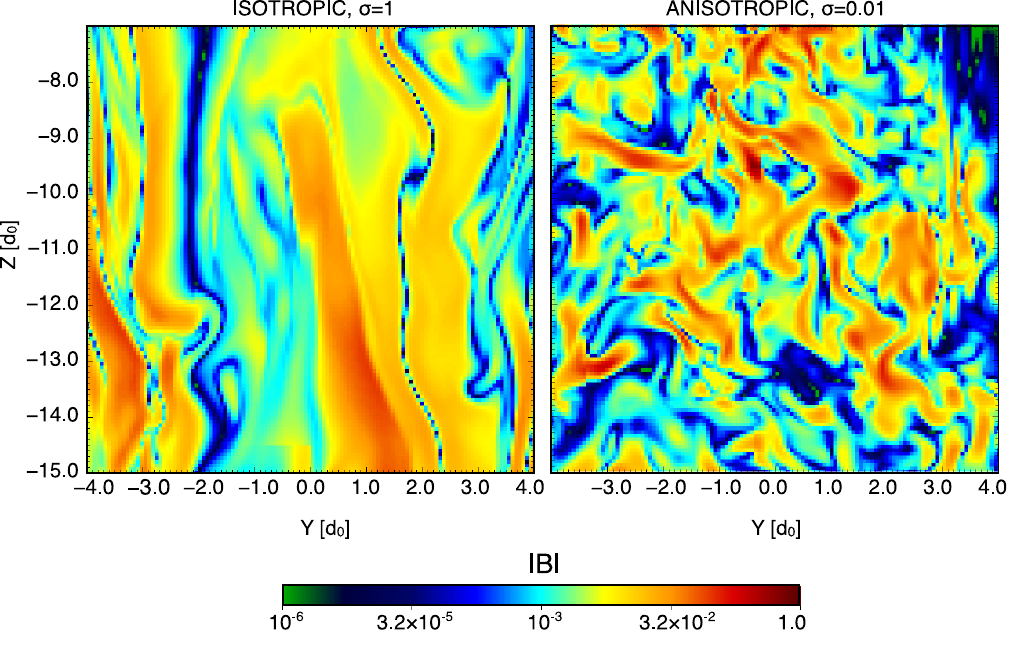}
    \caption{Details 
 of a 2D slice of the magnetic field module (in code units) in a portion of the tail of two different BSPWNe \citep{Olmi_Bucciantini_2019_1}. The~pulsar is located at  (0, 0, 0)$\,d_0$. The~figure directly compares a low turbulent tail (left side) and a high turbulent one (right side). It is evident that, when turbulence is low, even far away from the injection, the~shocked wind shows a rather ordered structure, while when the level of turbulence is high, the~properties of the wind are completely lost due to the high mixing of the~plasma.} 
    \label{fig:bspw_turb}
\end{figure}

\begin{figure}[H]
    \centering
	\includegraphics[width=0.7\textwidth]{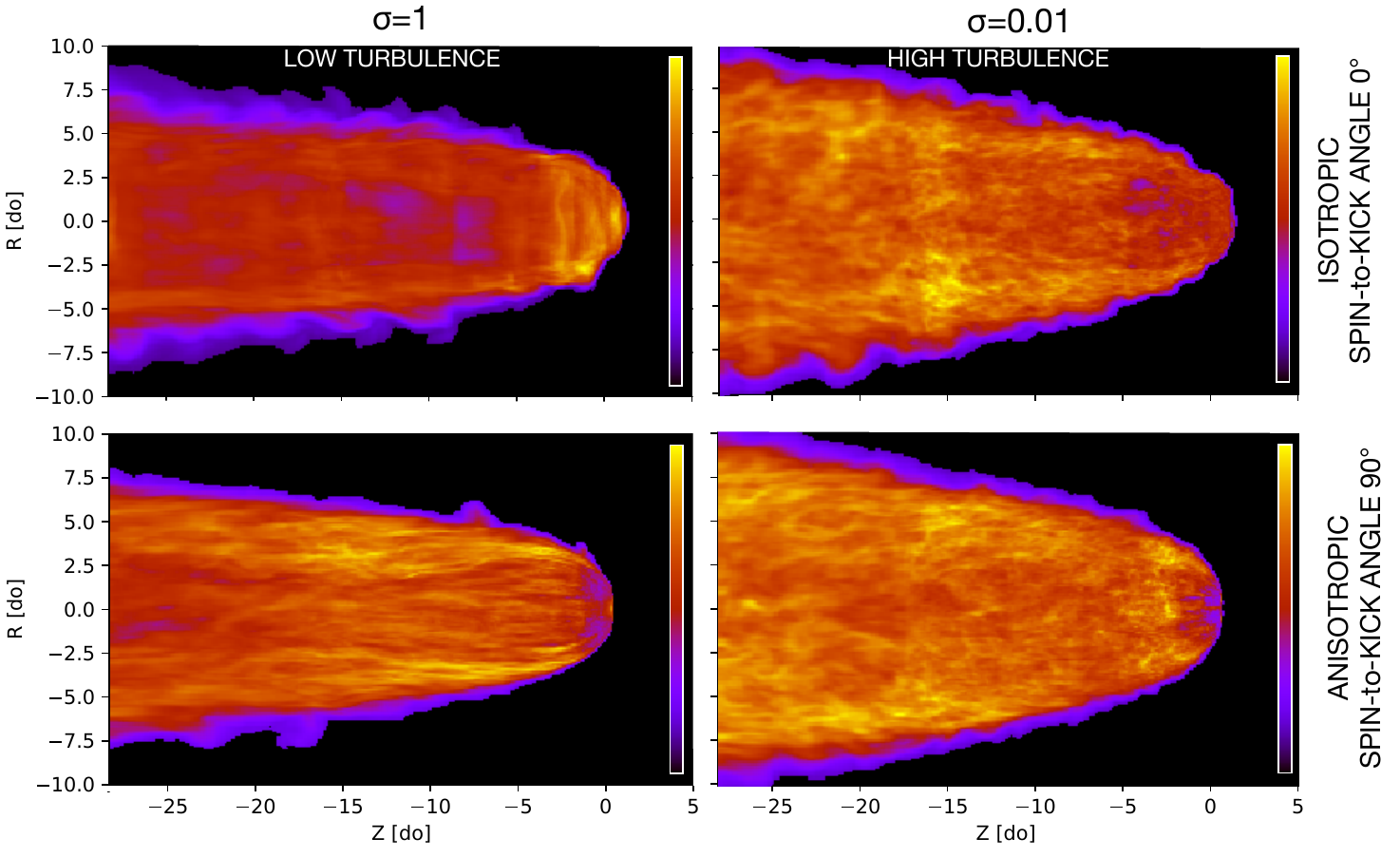}
    \caption{2D maps of the square root of the total synchrotron intensity, normalized to the maximum, computed as in \citet{Olmi_Bucciantini_2019_2}, comparing different systems characterized by a low turbulence level (panels on the left, with~high wind magnetization, $\sigma$) and a high turbulence one (panels on the right, with~low $\sigma$). It can be noticed how, when the system is dominated by turbulence, the~resulting intensity is rather uniform and it gives poor information about the properties of the wind. On~the contrary, when turbulence is low, intensity maps show diverse morphologies and can then give information about the physical properties of the emitting plasma. The~impact of turbulence on the polarization and the comparison of many other configurations can be found in  \citet{Olmi_Bucciantini_2019_2}. }
    \label{fig:compare-turb}
\end{figure}

BSPWNe simulations have been used to investigate a huge number of different situations: various geometrical configurations for the pulsar magnetic to spin axes inclination, as~well as diverse inclinations of the pulsar direction of motion with respect to the first two; different levels of magnetization and anisotropy of the pulsar wind; the interaction with a magnetized outer medium \citep{Barkov:2019,Olmi_Bucciantini_2019_1}.
These models show that the global dynamics of the bow shock is rather independent of the geometry of the system and on the wind properties. The~larger effect is registered when the inclination between the pulsar spin-axis and the star direction of motion is $\sim$45$^\circ$, showing up as perturbations of the forward shock (FS).
On the other hand, the~internal dynamics of the tail is largely dominated by the amount of magnetization and anisotropy that characterizes the pulsar wind. In~particular, regardless of the level of anisotropy, when the magnetization is low,  turbulence easily develops in the tail, down to small scales.
At odds with the magnetization, when the anisotropy increases, it also increases the level of turbulence \citep{Olmi_Bucciantini_2019_2}.
This behavior can be easily seen in Figure~\ref{fig:bspw_turb}, where we can directly compare a low turbulent system (high magnetization/low anisotropy) with a high turbulent one (low magnetization/high anisotropy).
Turbulence does not produce any sizable magnetic amplification, since the magnetic field tends to reach equipartition with the turbulent kinetic energy. 
However, the level of turbulence is very important for the persistence of the injection properties of the wind along the tail that, as~we will discuss in the following section, is a fundamental aspect for the possibility to have an efficient release of particles into the ambient.
Moreover, it is also relevant for the observational properties of the BSPWN; when turbulence is low there is an evident correlation between the pulsar wind injection conditions and the surface brightness of the source, with~different geometries reflecting in a variety of emitting morphologies (bright heads and faint tails, bright tails with fainter heads, bright lateral wings). On~the other side, when turbulence starts to dominate, all systems tend to be very similar in emission, making it harder to distinguish between different physical properties at injection (see Figure 
\ref{fig:compare-turb}).
Turbulence also has a direct impact on the polarization properties, with~the polarization fraction increasing with the tail magnetization \citep{Olmi_Bucciantini_2019_2}. 
%

%

\section{Particles Escaping from~BSPWNe}
\label{sec:esc}
As mentioned before, the~idea that energetic particles can escape the bow shock of a PWN was originally theorized by \citet{Bandiera:2008}, with~the aim to interpret the misaligned X-ray feature observed in the Guitar nebula.
In that work, the~author claimed that particles with a Lorentz factor close to that of the maximum potential drop ($\gamma\rs{MPD}$) are able to escape the bow shock and, once outside, to~stream along the ordered magnetic field lines of the ambient medium. 
To explain the observed properties, in~particular the orthogonal extension of the feature---its thickness, as~due to synchrotron radiation, the~magnetic field in the feature needs to be sufficiently large, \citet{Bandiera:2008} estimated for the Guitar's feature a field of $\sim$40 $\upmu$G,  and~similar values have been later inferred for other objects showing misaligned features \citep{Pavan:2014, Bordas:2020, deVries_Romani:2020, deVries_Romani:2022}.
As we will discuss in the following, how to explain such a large magnetic field in coincidence with the escaped particles, starting from the ambient magnetic field of order of few $\upmu$G, is one of the main difficulties in understanding the nature of this fascinating features.
{Particles in these X-ray features also show a harder photon index ($\Gamma_X$ $\sim$ 1--1.6) than the one usually measured in PWNe, possibly resulting from re-acceleration processes happening between the TS of the pulsar wind and the bow shock, thanks to the presence of converging flows \citep{Bykov:2017}.}

The process through which particles escape from the bow shock was later demonstrated thanks to numerical models. 
\citet{Bucciantini:2018} showed that current sheets and current layers are able to confine particles from their injection zone at the pulsar wind TS and, depending on the geometry, can easily carry them at the contact discontinuity with the ambient medium. 
This mechanism is also able to produce strong charge anisotropy in the escaping~flow.

Soon after, \citet{Barkov:2019} showed that  the appearance and persistence of reconnecting regions at the bow shock surface is common in 3D MHD models of BSPWNe immersed in an ordered magnetized medium. 
Particles transported by currents can then access the reconnecting point, depending on the geometry of the system, and~jump outside to stream away along the ISM magnetic field lines \citep{Olmi_Bucciantini_2019_3}. 
These 3D high-resolution MHD simulations permit one to compute the properties of the particle transport in the BSPWN. 
A huge number of electrons and positrons are injected along the termination shock of the pulsar wind, considering different energies. 
As expected, particles injected close to a current sheet remain easily trapped in it, and~can be collimated towards a reconnection~point. 

The efficiency of the escaping process is energy dependent: what matters for the escape is in fact the ratio of the particle Larmor radius in the BSPWN magnetic filed (at the head, $B\rs{bs}$) $R_L= m c^2 \gamma/(e B\rs{bs})$ to the characteristic size of the bow shock, $d_0$ \citep{Bucciantini:2018}.
Equating $R_L = d_0$, we found a relation between the Lorentz factor of the escaping particle ($\gamma\rs{esc}$) and the Lorentz factor associated with the maximum potential drop:
\begin{equation}\label{eq:energydepesc}
    \gamma\rs{esc} \!\!\!= \gamma\rs{MPD}\,\left(\frac{B\rs{bs}}{B_0}\right)\,\left(\frac{v_A}{\vpsr}\right) \!\!\simeq 0.5 \,\gmpd\,
    \left(\frac{B\rs{bs}}{50 \, \upmu\mathrm{G}}\right)\,
    \left(\frac{B_0}{3 \, \upmu\mathrm{G}}\right)^{-1} \left(\frac{v_A}{10 \,\mathrm{km/s}}\right) \,
    \left(\frac{\vpsr}{350 \,\mathrm{km/s}}\right)^{-1},
\end{equation}
where $v_A = B_0/\sqrt{4\pi \rhoism}$ is the Alfvèn velocity in the ISM and all the quantities have been expressed with characteristic scalings for the considered problem. 
Particles can easily escape if $R_L \gtrsim d_0$, meaning that 
 $\gamma\rs{esc} \gtrsim \gamma\rs{MPD}\, (B\rs{bs}/B_0)(v_A /\vpsr)$.
Only those particles carrying a relevant fraction of the maximum available energy in the system ($\gtrsim$50\% $\gmpd$), corresponding for standard values (see Equation~(\ref{eq:maxpd})) to \mbox{50--100 TeV}, efficiently escape from the bow shock head, while lower energy ones have a marginal probability to leave the system.
Long living low energy particles more likely escape from the less bounded back part of the bow shock tail, far away from the pulsar, provided that they can survive long enough against synchrotron~losses.

As shown in Figure~\ref{fig:bspw_esc}, the~{higher} the energy of the particles (thus the {larger} their gyration radius), the~more diffuse is their escaping process.
On the contrary, particles of lower energies (but still above the threshold for an efficient escape) easily form collimated jet-like features.
One of the most important geometrical aspects to determine the appearance of the feature in the ISM is the inclination of the pulsar spin-axis with respect to the direction of the ambient magnetic filed. The~inclination of the spin-axis in fact determines the direction to which the equatorial and polar current sheets of the pulsar wind point (see Figure~\ref{fig:bspw_esc}), then the regions where particles are conveyed towards the magnetopause. 
This not only easily gives very asymmetric features, but~also produces charge separation of the escaping particles, confirming what was previously found with a much simpler modeling by \citet{Bucciantini:2018}.
As we will discuss in the following, charge separation is a key ingredient for the amplification of the ambient magnetic field, thanks to the formation of a net electric current in the~ISM.
\vspace{-6pt}

{\begin{figure}[H]
    \centering
	\includegraphics[width=.7\textwidth]{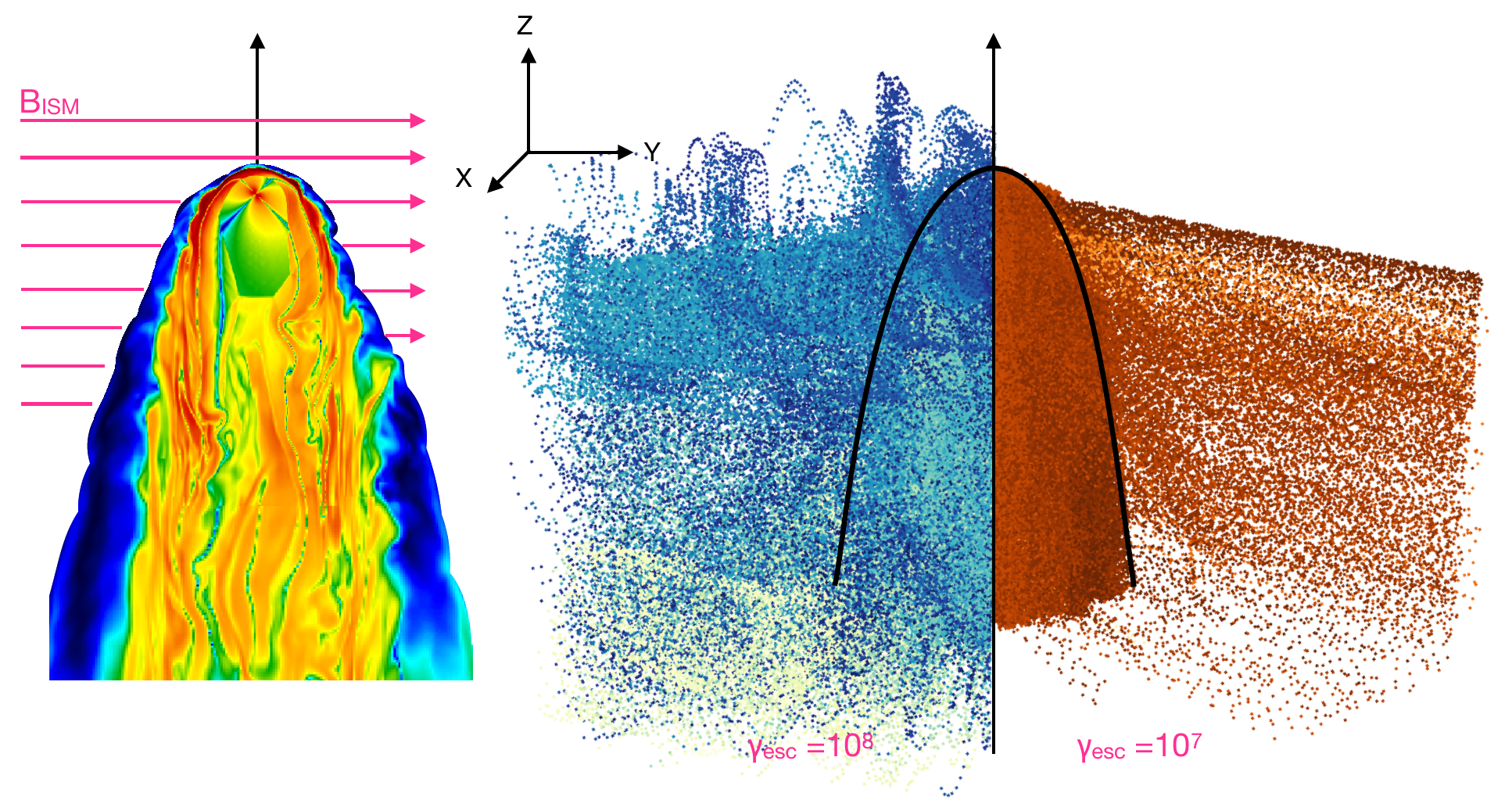}
    \caption{Left image: 2D slice of the magnetic field intensity from a 3D MHD simulation of a BSPWN \textls[15]{from \citet{Olmi_Bucciantini_2019_3}. The~pulsar motion is directed along the $Z$ axis, while the ambient} } 
    
        \label{fig:bspw_esc}
\end{figure}
    
    \captionof*{figure}{magnetic field (uniform) is directed along $Y$. In~this case, the pulsar spin axis is inclined by $45^\circ$ with respect to the direction of motion, the~magnetic filed by $90^\circ$. This configuration maximizes the formation of asymmetric features thanks to the distribution of the current sheets (visible as four orthogonal thin black lines forming at the PSR location, inside the TS).
    Right image: composite projection of a 3D representation of the particles escaping from the same BPWNe on the left side. The~border of the bow shock is drawn with a black line. The~leftmost half of the plot (blue colored) shows particles with a Lorentz factor of $\gamma=10^8$, escaping in a rather diffusive way, while lower energy ones (rightmost half---red colored) tend to escape in a collimated feature.} }
\vspace{6pt}

\section{Outside of the~Source}
\label{sec:out}
Although we are confident that we have understood the main aspects of the mechanisms that allow energetic particles to escape from an evolved PWN, we still lack an understanding of what happens outside the source.
As mentioned before, to~explain the observed emission from the misaligned X-ray features formed by BSPWNe, a~magnetic field of order of a few tens of $\upmu$G must be assumed; the ambient magnetic field needs then to be amplified by a factor $\sim$10.
On the other side, the magnetic field in TeV halos seems to be lower than the ambient one; deep X-ray studies in fact failed to detect any synchrotron radiation from the halo surrounding Geminga, which one expects if the magnetic field there is of the order of a few $\upmu$G, and~this sets an upper limit of 1 $\upmu$G \citep{Liu:2020}.
Moreover, the observed extension of these halos suggested a high suppression of the diffusion length, with~a diffusion coefficient $\sim$2 orders of magnitude below the average galactic one. The~physical conditions of the ambient medium in the sources surroundings then appear strongly modified with respect to the standard~ISM.

The nature of TeV halos has been largely debated recently, and~a number of different interpretations proposed \citep{Lopez-coto:2022}.
A very interesting one is that they arise from the modifications that escaping electrons and positrons produce in the ISM while leaving their parent source \citep{Evoli:2018}. Escaped particles can in fact  confine themselves through the amplification of magnetic field fluctuations via the resonant streaming instability.
But to efficiently amplify the magnetic field, a~sort of fine tuning to rather uncomfortable values of the parameters is necessary in this~interpretation.

A more efficient way to amplify the magnetic field is instead through the non-resonant streaming instability, also known as the Bell's instability \citep{Bell:2004}.
To be triggered, the~instability essentially requires that two conditions are~realized:
\begin{itemize}
 \item An electric current is present;
 \item The energy density carried by the escaping particles is larger than the magnetic energy density of the ambient medium.
\end{itemize}

\textls[-30]{The~first point is directly satisfied according to the results \mbox{by \citet{Olmi_Bucciantini_2019_3}}}, since charge separation of the escaping leptons means that an electric current is likely to develop as soon as particles run out of the source.
If $n\rs{esc}$ is the number density of the escaping particles, the~second condition can be written as:
\begin{equation}\label{eq:satcond}
    n\rs{esc} m c^2 (\gamma\rs{esc} -1) \gtrsim \frac{B_0^2}{8\pi}\,.
\end{equation}
For 
 simplicity, here we assumed that all the particles escape with the same Lorentz factor ($\gamma\rs{esc}$), while we might expect a range of energies distributed following the particles injection spectrum.
The number of particles escaping to form the current can be linked to the pulsar spin-down power through the escaping efficiency ($\psi\rs{esc}$).
It is possible to show that the previous condition can be satisfied if $\psi\rs{esc}\gtrsim 10^{-6}$, no matter the distribution of the particles pitch angles at their injection into the outer~medium.

Such a fraction does not appear to be a problem, since the pulsar injects a much larger amount of its spin-down energy into particles (up to a few tens of per cent), so  Bell's instability can be theoretically excited and cause the modifications to the ambient field needed to explain observations.

In the case of halos, the energy density of particles in the source vicinity can be estimated from gamma-ray observations, and~it results in an order of $10^{-2}$ eV/cm$^3$ \citep{Giacinti:2020}. If~the magnetic field in the halo is suppressed to values below 1 $\upmu$G, the~energy density associated with the magnetic field can easily go below this value, satisfying the previous condition number two.
The same direct check with misaligned X-ray features is more complicated. 
If one estimates the energy density stored in the feature from the observed X-ray luminosity, assuming an injection time corresponding to the time the pulsar needs to cross the apparent orthogonal dimension of the feature (its thickness), finds a value of few $\times 10^{-2}$ eV/cm$^{-3}$, about one order of magnitude below the magnetic energy density associated with the $\sim$3 $\upmu$G ISM field. 
However, this estimation is much too simplified: first of all, it is likely that not all the escaped particles will contribute to the emission that we can detect, but~only those pointing towards us in their motion.
There is then no real motivation to assume that only particles with an energy above the threshold for the X-ray emission escape from the bow shock: we might then miss part of them in our observations.
Finally, we have no idea of the real geometry of the escaping flow, and~it might differ substantially from a jet-like~one.

To date, it seems that no superposition of the two kind of manifestations of particles escape exists. None of the observed systems show an extended X-ray misaligned tail associated to a diffused TeV halo.
Of course, given the still limited number of sources firmly detected in both categories, it is not possible to exclude that the two features can appear simultaneously.
If we did not find a coincidence of the two, what might discriminate between the formation of a misaligned tail or a halo are the properties of the ambient medium more than the properties of the escaping particles.
If the turbulence generated by the instability is sufficient to explain the suppression of the diffusion length in the halos, or~if it can be sustained long enough to amplify the magnetic field to allow the formation of the thin X-ray jets is still a matter of investigation (Bandiera, Olmi~et~al., in~preparation).

\section{Concluding~Remarks}
\label{sec:concl}
X-ray thin, extended, and strongly misaligned tails, or~jet-like features, have been detected in the last years in an increasing number of evolved systems.
In principle, they can be a rather common manifestation, and~the still limited number of known objects might be biased by our limited capability to find them, combined with geometrical motivations. Detecting such features is in fact not an easy task: their emission is faint and in general can be extracted from the background only with very deep observations. 
Together with the detection of extended TeV halos surrounding evolved pulsars, these objects represent a smoking gun for the efficient escape of charged particles from evolved systems.
Evolved pulsar and their nebulae are then certainly fundamental sources of cosmic rays in our Galaxy and, possibly, among~the primary sources of the measured positron excess in their spectrum at~Earth.

Here we showed that many of their observational properties can nowadays be understood thanks to 3D relativistic MHD models.
They allowed us to demonstrate how particles can flow outside of these sources, thanks to the combined effect of the survival of current sheets in the shocked pulsar wind, inside the nebula, and~the formation of reconnection points at the magnetopause separating the pulsar wind nebula from the outer medium.
The development of a low level of turbulence is fundamental for the maintenance of the coherent structure of the current sheets from the injection (at the pulsar wind termination shock) to the bow shock PWN tail. This puts constraints on the physical properties of the pulsar wind (its magnetization level and anisotropy).
We also proved that the escaping process is energy dependent: only particles carrying a large fraction of the maximum available energy in the system ($\gtrsim$50\% $\gmpd$) can efficiently escape. For~common values of the parameters describing these systems, this means that only particles with energies above tens to a hundred of TeV flow out from the source.
{As largely discussed in the past, besides~being interesting by themselves, PWN also provide indirect evidence for the physics of the pulsar that fuels the nebula \citep{Bejger:2003}. The~presence of escaping particles that implies that pulsars can accelerate particles very close to the theoretical limit is a new fundamental step in the our understanding of these stars.}

The escaping flux is also characterized by a certain level of charge separation. This is a very important ingredient to make possible the growth of the non-resonant streaming instability (Bell's instability), that might lead to the modification of the ISM properties in the vicinity of those sources, allowing for the formation of TeV halos (possibly developing in regions where the ambient magnetic field is low from the beginning) or misaligned, collimated X-ray tails (where the standard $\sim$3 $\upmu$G magnetic field can be efficiently amplified by the onset of the instability).

What is still missing is to firmly confirm that this mechanism can work long enough to produce the observed features. This will require further theoretical investigations (ongoing) and numerical efforts to investigate the particle transport and properties of the instability they can excite in the regime characteristic of the present problem.
Understanding these processes is extremely important for assessing what happens to energetic particles once they escape from their parent  sources, and~how they interact with the ambient medium. This has direct consequences for our understanding of cosmic ray propagation in the Galaxy and how it is modified in the surroundings of a source.
Moreover, understanding the physical processes leading to the formation of halos or misaligned tails and how widespread these manifestations are is also relevant for the interpretation of present and future gamma-ray data, where the number of expected contributions from evolved PWNe is~large.

 \vspace{6pt}
\funding{This research was funded by  the Italian Space Agency (ASI) and National Institute for Astrophysics (INAF) under the agreements ASI-INAF n.2017-14-H.0, from~INAF under grants:  ``INAF Mainstream 2018 \textit{Particle Acceleration in Galactic Sources in the CTA era}'', ``PRIN-INAF 2019 \textit{From massive stars to supernovae and supernova remnants: driving mass, energy and cosmic rays in our Galaxy}'' and ``INAF MiniGrant \textit{Numerical Studies of Pulsar Wind Nebulae in The Light of IXPE}''.
}


\acknowledgments{The author wishes to thank N. Bucciantini and R. Bandiera for the continuous collaboration on the argument treated in this manuscript. She also acknowledges stimulating discussions with P. Blasi and E.~Amato.}

\conflictsofinterest{The author declares no conflict of interest. The funders had no role in the design of the study; in the collection, analyses, or interpretation of data; in the writing of the manuscript; or in the decision to publish the~results.
} 

\abbreviations{Abbreviations}{
The following abbreviations are used in this manuscript:\\
 
 \vspace{-6pt}
\noindent 
\begin{tabular}{@{}ll}
BSPWN & Bow Shock Pulsar Wind Nebula \\
HD & HydroDynamic \\
IC & Inverse Compton  \\
ISM & InterStellar Medium \\
\end{tabular}
}

\noindent 
\begin{tabular}{@{}ll}
MHD & ~~~~Magneto-HydroDynamic \\
PSR & ~~~~Pulsar \\
PWN & ~~~~Pulsar Wind Nebula  \\
SN & ~~~~Supernova \\
SNR & ~~~~Supernova Remnant  \\
\end{tabular}




\reftitle{References}



\end{paracol}


\end{document}